\documentclass[aps,prd,english,preprintnumbers,amsmath,amssymb,nofootinbib,twocolumn,superscriptaddress]{revtex4-1}
\usepackage[T1]{fontenc}
\pdfoutput=1

\usepackage{graphicx}
\usepackage{color}
\usepackage{bbm}
\usepackage{amssymb}
\usepackage{amsmath}
\usepackage{tabularx}
\usepackage{dcolumn}

\usepackage{dsfont}

\def\0#1#2{\frac{#1}{#2}}

\def\s0#1#2{\mbox{\small{$ \frac{#1}{#2} $}}}

\newcommand{\beq}{\begin{equation}}
\newcommand{\eeq}{\end{equation}}
\newcommand{\bea}{\begin{eqnarray}}
\newcommand{\eea}{\end{eqnarray}}

\usepackage{babel}
\makeatother
\begin{document}

\title{Polarized fermions in one dimension: density and polarization \\ from complex Langevin calculations,
perturbation theory, and the virial expansion}

\author{Andrew C. Loheac}
\email{loheac@live.unc.edu}
\affiliation{Department of Physics and Astronomy, University of North Carolina, Chapel Hill, North Carolina 27599, USA}

\author{Jens Braun}
\email{jens.braun@physik.tu-darmstadt.de}
\affiliation{Institut f\"ur Kernphysik (Theoriezentrum), Technische Universit\"at Darmstadt, 
D-64289 Darmstadt, Germany}
\affiliation{ExtreMe Matter Institute EMMI, GSI, Planckstra{\ss}e 1, D-64291 Darmstadt, Germany}

\author{Joaqu\'in E. Drut}
\email{drut@email.unc.edu}
\affiliation{Department of Physics and Astronomy, University of North Carolina, Chapel Hill, North Carolina 27599, USA}

\begin{abstract}
We calculate the finite-temperature density and polarization equations of state of one-dimensional fermions with a zero-range 
interaction, considering both attractive and repulsive regimes. {In the path-integral formulation of the grand-canonical ensemble, 
a finite chemical potential asymmetry makes these systems intractable for 
standard Monte Carlo approaches due to the sign problem. 
Although the latter can be removed in one spatial dimension, 
we consider the one-dimensional situation in the present work to provide an efficient test for studies of the higher-dimensional
counterparts. To overcome the sign problem, we use the complex Langevin approach, which we compare here with 
other approaches: imaginary-polarization studies, third-order perturbation} theory, and the third-order virial expansion. 
We find very good qualitative and quantitative agreement across all methods in the regimes studied, which supports their validity.
\end{abstract}

\maketitle

\section{Introduction}

Motivated by the potential appearance of exotic polarized superfluid phases in ultracold atoms
(see Refs.~\cite{2004RvMP...76..263C, 2010RPPh...73g6501R, 2010RPPh...73k2401C, GUBBELS2013255} for reviews), along with the possibility
of importing powerful methods from relativistic lattice field theory to the area of nonrelativistic strongly correlated matter
(see e.g.~\cite{Porter:2016vry, Rammelmuller:2017myk,Drut:2017fsv,Loheac:2017uuj, Shill:2017jtm, Loheac:2017yar}), 
we report on the determination of the thermal properties of one-dimensional (1D) fermionic systems at finite 
chemical potential asymmetry, i.e. polarized fermions. While 1D fermions have been extensively studied (see e.g. 
Refs.~\cite{TakahashiBook, Giamarchi, RevModPhys.85.1633}), we use them here as a testbed for a suite of methods that are 
applicable to {their higher 
dimensional counterparts.}

Indeed, recently we applied complex stochastic quantization~\cite{Loheac:2017yar} and imaginary-polarization~\cite{Loheac:2015fxa}  
methods to the analysis of 1D fermions whose higher-dimensional analogues have a sign problem.
Such is the case, for instance, for Hamiltonians featuring repulsive interactions~\cite{Loheac:2017yar}, finite chemical 
potential asymmetry (i.e. finite polarization)~\cite{Braun:2012ww}, {finite mass 
imbalance~\cite{Braun:2014pka, Rammelmuller:2017vqn,Roscher:2013aqa}, or} both~\cite{Roscher:2013cma, Roscher:2015xha}.
In this work, we continue those investigations by tackling the polarized 1D Fermi gas with
both attractive and repulsive interactions, putting together a more diverse {set of tools than} in our previous work: we compare calculations
performed with the complex {Langevin approach (CL) with} {those obtained 
from hybrid Monte Carlo (MC) studies at imaginary polarization (iHMC), 
lattice} perturbation theory at third order (N3LO), and the virial expansion (at third order).

Our objective is to establish the reliability of non-perturbative approaches such as the CL method to then proceed
to problems in higher dimensions, such as the {spin-$1/2$ Fermi gas tuned to the unitary limit.}
While that and similar systems have been extensively studied in their unpolarized states, 
the polarized 3D case remains a mystery in many ways. There, 
the possible appearance of inhomogeneous superfluid phases at low temperatures has attracted a lot of attention in recent years 
(see, e.g., Refs.~\cite{2004RvMP...76..263C, GUBBELS2013255} for reviews). Still, what little is known about 
the fate of such phases in calculations beyond 
the mean-field approximation remains unclear at present (see, e.g., Refs.~\cite{Roscher:2015xha,2017arXiv170300161W,2018arXiv180403035F} for recent studies including fluctuation effects)
and calls for {\it ab initio} studies.  However, the latter (in form of MC methods) only allow investigations of 
unpolarized fermions with attractive contact interactions. The spin-polarized counterpart poses the aforementioned sign problem since the fermion determinant 
corresponding to each species may generally take different signs, producing a non-positive probability measure.

One way of avoiding non-positive probability measures is given {by the so-called iHMC method, whereby} one takes the chemical potential for each species to be 
complex, such that the chemical potential for one species is the complex conjugate of the other. {The product of these two fermion determinants is then positive definite 
and represents a valid probability measure. However, this comes at a price: the calculated observables now have to be analytically continued to the real axis to 
obtain the physical observables. That} technique was applied by the present authors to the 1D case of polarized, attractively 
interacting fermions in Ref.~\cite{Loheac:2015fxa} with success for moderate-strength couplings, but {the technique 
was found to be difficult to apply in the case of very strong couplings. Moreover, it} is limited to attractive interactions, and is cumbersome 
in the sense that an appropriate ansatz must be selected to fit the Monte Carlo results obtained on the imaginary axis. 

{The main objective of our present work is to provide further validations of our CL approach to non-relativistic Fermi gases rather 
than providing a detailed phenomenological discussion of the thermodynamics of one-dimensional Fermi gases. 
Against this background,
the remainder of this} paper is organized as follows. In Sec.~\ref{Sec:Methods} we review the path integral formalism leading to
the imaginary-polarization and complex Langevin methods, with emphasis on the latter; in Sec.~\ref{Sec:PT} we review
the perturbation theory formalism leading to our N3LO results; in Sec.~\ref{Sec:VE} we discuss the elements of the virial
expansion, which is non-perturbative and which we use to validate our results in the {low-fugacity region. Note 
that our discussion of the various methods is meant to be minimalistic as detailed discussions and introductions to the 
tools underlying our present work can be found in
Ref.~\cite{Lee:2008fa,Drut:2012md} regarding MC approaches to non-relativistic systems, Refs.~\cite{Braun:2012ww,Loheac:2015fxa} regarding iHMC,
and Ref.~\cite{Loheac:2017yar} regarding 
our perturbative approach. 
In Sec.~\ref{Sec:Results}, we present our results for the density and polarization equations of state, including a brief discussion of 
the underlying systematics. Finally, in} Sec.~\ref{Sec:Conclusions}, we
summarize and present our conclusions.

\section{Stochastic Methods \label{Sec:Methods}}

\subsection{Basic formalism}
As in most finite-temperature calculations, we choose the grand-canonical ensemble, where
the partition function is defined by
\beq
\mathcal Z = \text{Tr}\left[\exp(-\beta \hat K) \right]\,,
\eeq
where $\hat K = \hat H - \mu_\uparrow \hat N_\uparrow- \mu_\downarrow \hat N_\downarrow$ 
{and~$\uparrow, \downarrow$ refers to two particle species.}
Here, $\hat H$ is the Hamiltonian, $\beta$ is the inverse temperature, $\mu_s$ 
is the chemical potential for spin-$s$ particles, and $\hat N_s$ is the corresponding particle 
number operator. Below, we will also use the {notation
\bea
\mu \equiv (\mu_\uparrow +\mu_\downarrow)/2\,,\qquad 
h \equiv (\mu_\uparrow  - \mu_\downarrow)/2\,,
\eea
such that
\bea
\mu_\uparrow = \mu + h\,,\qquad 
\mu_\downarrow = \mu - h\,.
\eea
The} Hamiltonian we will use is of the standard form
\beq
\hat H = \hat T + \hat V,
\eeq
where $\hat T$ is the kinetic energy operator, and $\hat V$ is the potential energy operator
given by
\beq
\hat T = \int dx \;
\sum_{s=\uparrow,\downarrow}\! \hat \psi_s^\dagger({x})\!\left(\!-\frac{\hbar^2}{2m} \frac{d^2}{dx^2}\!\right)\! \hat \psi_s^{}({x})\,,
\eeq
and
\beq
\hat V = - g \int dx \; \hat n^{}_{\uparrow}({x}) \hat n^{}_{\downarrow}({x})\,,
\eeq
where $\hat \psi_s^{\dagger}, \hat \psi_s^{}$ are the creation and annihilation operators in coordinate space for particles of
spin $s$, and $\hat n^{}_{s} = \hat \psi_s^{\dagger}\hat \psi_s^{}$ are the corresponding density operators.

Below, we will put this problem on a spacetime lattice of spacing $\ell = 1$ in the spatial direction (which {sets the
scale for everything} else in the computation) and extent $L = N_x \ell$, and spacing $\tau$ in the imaginary-time 
direction, such that $\beta = \tau N_\tau$. Thus, $N_x$ and $N_\tau$ are the number of lattice points in the {spatial and 
time directions, respectively. We use periodic boundary conditions for the former, and anti-periodic for the latter in order to respect 
the statistics of the fermion fields.}

By applying a Suzuki-Trotter factorization first, one may use a Hubbard-Stratonovich (HS) transformation
to decouple the interaction, which comes at the price of introducing a field integral. We thus arrive at the starting point 
of many conventional methods used to compute thermodynamic observables, namely the field-integral representation 
of the grand-canonical partition function,
\beq
\mathcal{Z} = \int \mathcal{D}\sigma \, {\det} M_\uparrow[\sigma] \, {\det} M_\downarrow[\sigma]\,.
\eeq
Here, $M_s$ are the fermion matrices for each particle species (see Ref.~\cite{Loheac:2017yar} for details), and $\sigma$ is the auxiliary field introduced by our choice of HS transformation.
In most auxiliary-field MC methods, one then attempts to evaluate the integral stochastically by 
identifying a probability $P[\sigma]$ and corresponding action $S[\sigma]$ via
\beq
P[\sigma] = \exp(-S[\sigma]) = {\det} M_\uparrow[\sigma] \, {\det} M_\downarrow[\sigma]\,.
\eeq
{As a consequence, the calculation of observables takes} the form
\beq
\langle \mathcal O \rangle = \frac{1}{\mathcal Z} \int \mathcal D \sigma \, e^{-S[\sigma]} \mathcal O[\sigma]\,,
\eeq
such that the expectation value can be determined by sampling the auxiliary field $\sigma$ according to $P[\sigma]$.

\subsection{Imaginary polarization method}

As is well known, conventional MC algorithms are {usually not suitable for calculations 
at finite polarization because} $P[\sigma]$ is either 
complex or real but of varying sign, i.e. it suffers from the so-called phase or sign problem.
One way to guarantee a non-negative $P[\sigma]$ for systems with attractive interactions (where the 
sign problem comes from $\mu_\uparrow \neq \mu_\downarrow$) is to make the chemical potential
asymmetry $h$ imaginary, such that $\mu_\uparrow$ and $\mu_\downarrow$ (and therefore 
${\det} M_\uparrow[\sigma]$ and ${\det} M_\downarrow[\sigma]$) are complex conjugates of each {other. We then have}
\beq
P[\sigma] = | {\det} M_\uparrow[\sigma]|^2.
\eeq
Such an approach, referred to above as iHMC, 
enables non-perturbative calculations of observables which are {\it a posteriori} analytically continued to
real asymmetry, as was done for the systems considered here in Ref.~\cite{Loheac:2015fxa},
and for mass-imbalanced systems in Refs.~\cite{Braun:2014pka, Rammelmuller:2017vqn}.

\subsection{Complex Langevin method}

Another way to bypass or overcome the sign problem is the CL method, which we will briefly describe here
following our work of Ref.~\cite{Loheac:2017yar}. The first step in the CL approach is to complexify the 
auxiliary field $\sigma$, such that
\beq
\sigma = \sigma_R + i \sigma_I,
\eeq
where $\sigma_R$ and $\sigma_I$ are real fields. The CL equations of motion, including a regulating 
term which prevents uncontrolled excursions into the complex plane (see Ref.~\cite{Loheac:2017yar}), are
\bea
\delta \sigma_R &=& -\text{Re}\left[\frac{\delta S[\sigma]}{\delta \sigma}\right] \delta t - 2\xi\sigma_R \delta t+ \eta \sqrt{\delta t}, \\
\delta \sigma_I &=& -\text{Im}\left[\frac{\delta S[\sigma]}{\delta \sigma}\right] \delta t - 2\xi\sigma_I\delta t,
\eea
where $\eta$ is a $t$-dependent noise field that satisfies $\langle \eta(x,\tau) \rangle = 0$ and 
$\langle \eta(x,\tau) \eta(x',\tau') \rangle = 2 \delta_{x,x'}\delta_{\tau,\tau'}$, and $\xi$ is a real parameter 
which for the following results is {set to $\xi=0.1$, see Refs.~\cite{Loheac:2017yar,Rammelmuller:2017vqn} for an analysis of the dependence of 
physical results on this parameter.} 
Note that the time $t$ is a fictitious time that is unrelated to
the imaginary-time $\tau$. In the CL context, $S[\sigma]$ is interpreted as a complex function of 
the complex variable $\sigma$; note that in the unpolarized case with attractive interactions, 
$\sigma$ becomes a real field.

The conditions for the validity of the CL algorithm 
have been extensively explored in recent years (see e.g.~\cite{Aarts:2009uq,Aarts:2011ax,Aarts:2017vrv,Nagata:2016vkn}),
as the CL method is not always guaranteed to converge to the right answer (in contrast with 
conventional stochastic quantization based on real actions).
When CL does converge correctly, the expectation values of observables 
$\langle \mathcal O \rangle$ are obtained by averaging over the real part of
$\mathcal O[\sigma]$, with complex fields $\sigma$ sampled throughout the CL evolution. 

In the path toward making CL a viable solution to the sign problem, 
problems were identified affecting convergence and correctness; one of the
most important of such problems was the appearance of uncontrolled excursions of $\sigma$ into the complex plane.
This issue is currently under investigation and a few approaches have been proposed (see e.g.~\cite{Bloch:2017sfg, Bloch:2017ods}).
In our case, we modified the action in a way reminiscent of the dynamical stabilization approach of 
Ref.~\cite{Attanasio:2016mhc,Aarts:2016qhx}, which {was proposed independently in Ref.~\cite{Loheac:2017yar} for non-relativistic systems.}

\section{Lattice perturbation theory\label{Sec:PT}}

In this section we outline the relevant formalism for our perturbation theory results.
We carried out our perturbative lattice calculations by expanding the grand-canonical partition function 
$\mathcal Z = \exp{(\beta P V)}$, 
as {in Ref.~\cite{Loheac:2017yar}. There, we carried out perturbation theory
starting from the field-integral formulation of the problem. That} expansion gives us direct access to the
pressure $P$ as a function of $\beta \mu$ and $\beta h$. 
Numerical differentiation with respect to $\beta \mu$ and $\beta h$ yields the density and 
polarization equations of state, respectively. Our perturbative calculations include contributions up to N3LO
in the auxiliary field coupling $A^2  = e^{\tau g} - 1$, where $\tau$ is the temporal lattice spacing and $g$
is the lattice coupling. Thus, the expansion takes the form
\beq
\frac{\mathcal Z}{\mathcal Z_0} = 1 + A^2 \Delta_1 + A^4 \Delta_2 + A^6 \Delta_3 + \dots,
\eeq
where the functions $\Delta_n(\beta \mu, \beta h)$ represent the contribution at order N$n$LO and $\mathcal Z_0$ is the noninteracting 
result. To access the pressure at a given order in $A^2$, we expand $\ln\mathcal{Z}$ in a consistent fashion
such that, at third order,
\beq
\label{Eq:FinalPTP}
\frac{P}{P_0} = 1 + \frac{1}{\ln\mathcal{Z}_0}\left( A^2\zeta_1 + A^4\zeta_2 + A^6\zeta_3 \right),
\eeq
where
\bea
\zeta_1 &=&  \Delta_1, \\
\zeta_2 &=&  \Delta_2 - \frac{1}{2} \Delta_1^2,\\
\zeta_3 &=&  \Delta_3 -  \Delta_1 \Delta_2 + \frac{1}{3}\Delta_1^3.
\eea
In Ref.~\cite{Loheac:2017yar} we presented calculations up to N3LO for the unpolarized case; here we extend those to
the polarized system for both attractive and repulsive couplings. Note that if we were again to perform the analysis of $\mathcal{Z}$ to a particular order of $A$, but consider distinct determinants for each flavor, we would arrive at the same symmetry factors and diagrams as for the unpolarized case, but find that exactly half of the propagators are a function of $z_\uparrow$, and the remaining half are a function of $z_\downarrow$. This translates to modifying the corresponding sums over momenta such that they are invariant under exchange of spin-up and spin-down fermions, and considering all permutations of $z_\uparrow$ and $z_\downarrow$ across non-commuting propagators. Note that these extra considerations lead to a small increase in computational complexity when evaluating these diagrams, particularly as the number of loops involved grows.

\section{Virial expansion\label{Sec:VE}}

In addition to the stochastic and perturbative results previously discussed, we compare to the equation of state provided 
by the virial {expansion, i.e. an expansion in 
powers of the fugacity~$z = \exp({\beta \mu})$. For $\beta \mu \ll -1$, such an expansion is indeed expected to be valid.}
\begin{figure*}[t]
	\centering
	\includegraphics[width=\columnwidth]{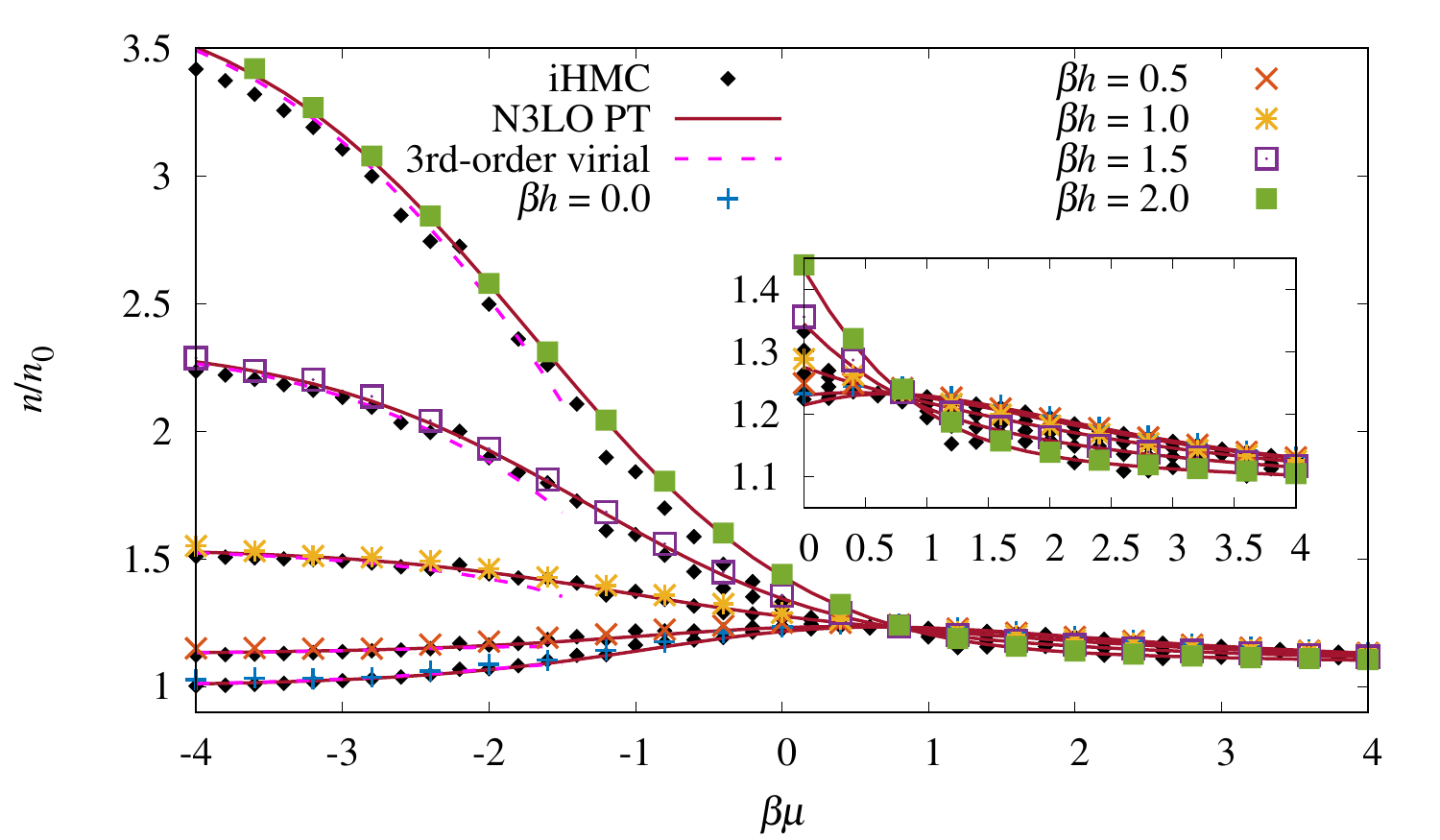}
	\includegraphics[width=\columnwidth]{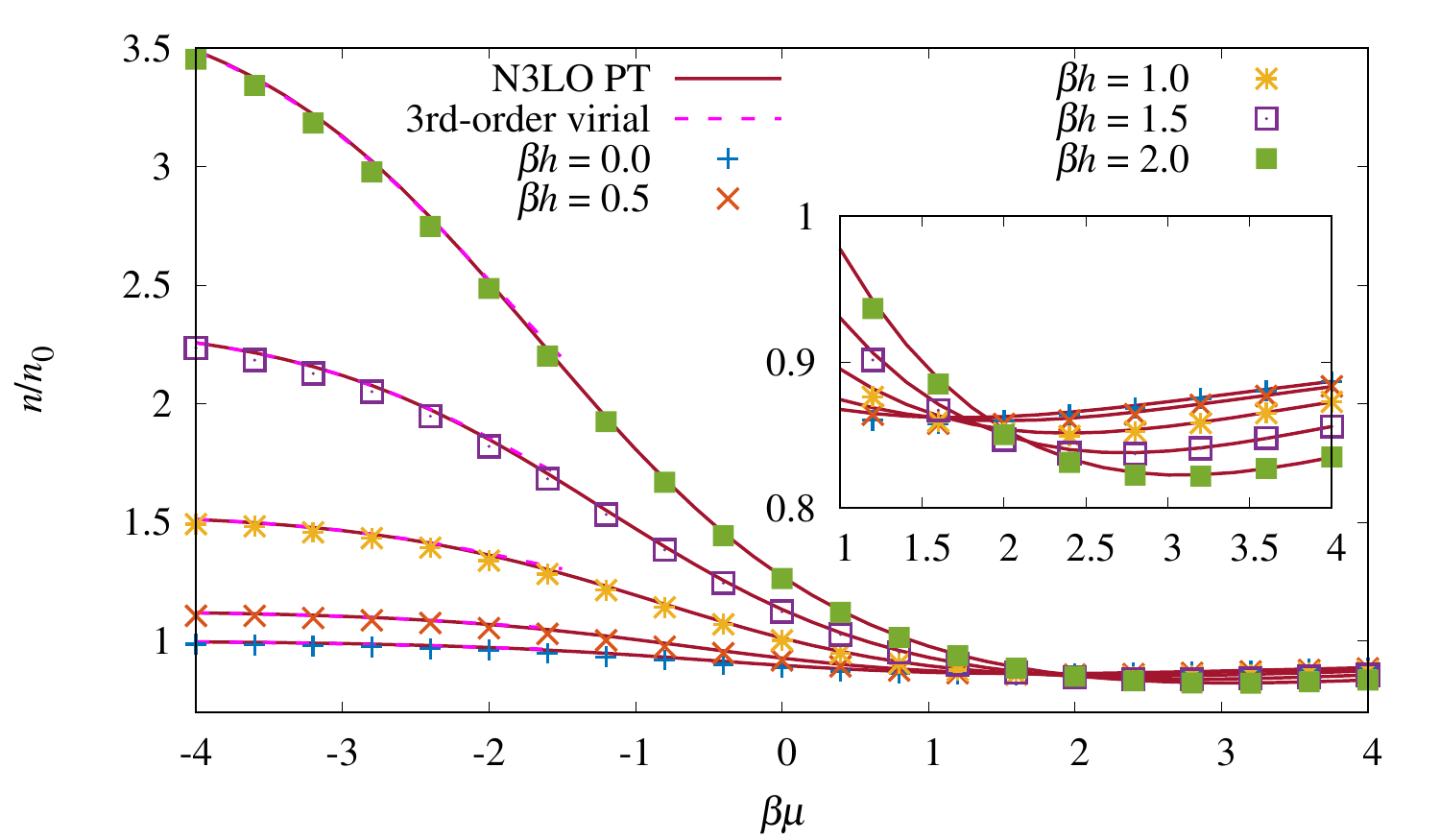}
	\caption{\label{Fig:DensityLambda1} Density equation of state $n = n_\uparrow + n_\downarrow$ normalized by the non-interacting, unpolarized counterpart $n_0$, 
	{for attractive (left) and repulsive (right) interactions of strength $\lambda = \pm 1$.
	Insets: Zoom in on the region $\beta\mu > 0$ (left) and $\beta\mu > 1$ (right).}
	In all cases, the CL results are shown with colored symbols, iHMC results (from Ref.~\cite{Loheac:2015fxa}) appear with
	black diamonds, perturbative results at third order are shown with solid lines, and virial expansion results
	appear as dashed lines.}
\end{figure*}

For unpolarized systems, the expansion reads
\beq
\ln \mathcal Z = -\beta \Omega = Q_1 \sum_{n=1}^{\infty}b_n z^n\,.
\eeq
where $\mathcal {Q}_1 = 2 L / \lambda_T$, $\lambda_T= \sqrt{2\pi \beta}$, and $b_n$ are {the virial coefficients.
The latter can} be obtained in terms of the $n$-particle canonical partition functions $Q_n$ using
\beq
\mathcal Z = \sum_{n = 0}^{\infty} Q_{n} z^n\,.
\eeq
For polarized systems, on the other hand, we write
\beq
\mathcal Z = \sum_{n,m = 0}^{\infty} Q_{n,m} z_\uparrow^n z_\downarrow^m\,.
\eeq
Note that $z_s = e^{\beta \mu_s}$ and, with our usual definitions, $\mu_\uparrow = \mu + h$ and $\mu_\downarrow = \mu - h$.

At leading order in $z_s$, we have $n^{}_{\uparrow,\downarrow}\lambda_T^{} = z^{}_{\uparrow,\downarrow}$, such that
\beq
 n\lambda_T^{} = (n^{}_\uparrow + n^{}_\downarrow)\lambda_T^{} = 2 {\rm e}^{\beta\mu}\cosh(\beta h)\,,
\eeq
which yields
\beq
\label{Eq:LOVirialn}
\frac{n}{n^{}_0} = \cosh(\beta h)\,.
\eeq
{Here, $n^{}_0$ is} the density for the unpolarized system; the above leading-order result holds for any interaction strength. 
Similarly, we find for the polarization that
\beq
\label{Eq:LOVirialm}
\frac{m}{n^{}_0} = \frac{n^{}_\uparrow - n^{}_\downarrow}{n^{}_0} =  \sinh(\beta h)\,,
\eeq
at leading {order in $z_s$.}

To access higher orders, we use the simpler expressions that result from taking the noninteracting case as a reference.
Thus, the usual unpolarized virial expansion of the pressure takes the form
\beq
-\beta \Delta \Omega =  \ln (\mathcal Z /\mathcal Z_0) = Q_1 \sum_{n = 2}^{\infty} \Delta b_n z^n,
\eeq
where $\Delta b_n = b_n - b_n^{(0)}$ is the change in the $n$-th order virial coefficient due to interactions.
{Note that the sum starts at $n=2$ since $b_1 \equiv b_1^{(0)} = 1$ by definition.}

For polarized systems, we have
\beq
-\beta \Delta \Omega =  \ln (\mathcal Z /\mathcal Z_0) = Q_1 \sum_{n,m = 1}^{\infty} \Delta b_{n,m} z_\uparrow^n z_\downarrow^m.
\eeq
Writing down the partition function in terms of the $(n,m)$-particle canonical partition functions $Q_{n,m}$, it 
{is straightforward to} see that
\bea
\Delta b_{1,1} &=& \Delta b_2, \\
\Delta b_{2,1} &=& \Delta b_{1,2} = \frac{\Delta b_3}{2},
\eea
which yields the first two terms of the virial expansion for the polarized case entirely in terms of the unpolarized {coefficients.

Differentiating} with respect to $z_s$ and dividing by the system size $L$ gives us access to $\Delta n = \Delta(n_\uparrow + n_\downarrow)$ and 
$\Delta m = \Delta (n_\uparrow - n_\downarrow)$. 
Using the relevant noninteracting polarized results $n^{(0)}$ and $m^{(0)}$, we can obtain $n$ and $m$ themselves. 
Calling $\bar n^{(0)}$ the noninteracting {\it unpolarized} result (i.e. $\bar n^{(0)} = \left . n^{(0)} \right |_{z_\uparrow = z_\downarrow}$), 
we have up to third {order,
\bea
\frac{n}{\bar n^{(0)}} &=& 
\frac{Q_1}{\bar n^{(0)} V} \! \left [ 2 \Delta b_2  z_\uparrow z_\downarrow \right. \nonumber \\
&& \qquad\quad \left. +  3\frac{\Delta b_3}{2} (z^2_\uparrow z_\downarrow+ z_\uparrow z^2_\downarrow) \right ]\!  +\!  \frac{n^{(0)}}{\bar n^{(0)}}\,.
\eea
Similarly, up to third order for the magnetization, we find}
\bea
\frac{m}{\bar n^{(0)}} = 
\frac{Q_1}{\bar n^{(0)} V} \left [ \Delta b_3 (z^2_\uparrow z_\downarrow - z_\uparrow z^2_\downarrow) \right ] + \frac{m^{(0)}}{\bar n^{(0)}}\,.
\eea
For reference, {we also present here the result for the density} and polarization of the polarized noninteracting Fermi gas:
\bea
n^{(0)} &=& \frac{1}{\sqrt{\pi} \lambda^{}_T } [I^{}_1(z_\uparrow) + I^{}_1(z_\downarrow)]\,,
\\
m^{(0)} &=& \frac{1}{\sqrt{\pi} \lambda^{}_T } [I^{}_1(z_\uparrow) - I^{}_1(z_\downarrow)]\,,
\eea
where $\lambda_T= \sqrt{2 \pi \beta}$,  $I^{}_1(z) = z\,{d I^{}_0(z)}/{d z}$, and 
\beq
I^{}_0(z) = \int_{-\infty}^{\infty}dx \ln(1 + z e^{-x^2})\,.
\eeq
{
In our numerical studies below, we employ the expressions for the density and the magnetization presented here at third 
order in the virial expansion, with the coefficients~$b_2$ and~$b_3$ taken from 
a numerical calculation~\cite{ShillInPreparation}, see also Tab.~\ref{Table:VirialCoefficients}.}
\begin{table}[t]
\begin{center}
\caption{\label{Table:VirialCoefficients}
Second and third-order virial coefficients $b^{}_2$ and $b^{}_3$
as a function of the dimensionless coupling $\lambda$. 
For the non-interacting gas ($\lambda=0$), the virial coefficients are $b^{}_n = (-1)^{n+1} n^{-3/2}$.
{At finite coupling, the interacting virial coefficients have been taken from a numerical calculation~\cite{ShillInPreparation}. 
The given values of~$b_2$ and~$b_3$ at~$\lambda=0$ correspond to the exact values.}
}
\begin{tabularx}{\columnwidth}{@{\extracolsep{\fill}}c|c|c}
$\lambda$ & $b^{}_2$ & $b^{}_3$ \\
\hline\hline
-2.0 & $-0.180(5)$  & $-0.0739(5)$ \\
-1.0 & $-0.490(5)$  & $0.394(5)$ \\
0 & $-0.35355\dots$  & $0.19245\dots$ \\ 
1.0 & $-0.0375(5)$  & $-0.240(5)$\\
2.0 & $0.190(5)$  & $-0.0615(5)$\\
\hline\hline
\end{tabularx}
\end{center}
\end{table}
%

\section{Results\label{Sec:Results}}

In this section we show our results for the density and polarization equations of {state as obtained from 
a non-perturbative calculation with the CL method on lattices of size up to $N_x = 61$ and $N_\tau = 160$, 
lattice perturbation theory up to N3LO using a matching lattice size, and 
the third-order virial expansion. In} addition, for attractive interactions we have at our disposal the data of Ref.~\cite{Loheac:2015fxa}, which were obtained using
the technique of imaginary polarization and analytic continuation (described above as iHMC).
The lattice calculations were performed using a temporal lattice spacing of $\tau = 0.05$ such that $\beta = \tau N_\tau$,\footnote{For a discussion of the dependence 
on the various parameters defining our space-time lattice, we refer the reader to Refs.~\cite{Loheac:2017yar,Rammelmuller:2017stp,Rammelmuller:2017vqn}.} 
which was chosen to provide a suitable balance between computational demand and finite-lattice effects. The CL calculations were performed using an adaptive Euler integrator, and were evolved for a total of $10^5$ iterations, where the first 10\% of samples were discarded to thermalize the system and improve convergence properties. 
Note that here we display the equation of state for a dimensionless coupling {strength at $|\lambda| = 1$ (i.e. for the repulsive 
and attractive case), but} additionally show results at $|\lambda| = 2$ in the Appendix.

\subsection{Density at $|\lambda|=1$}

In Fig.~\ref{Fig:DensityLambda1} we show our results for the density {equation of state at $\lambda = 1$ (left) 
and $\lambda = -1$ (right), as} a function of $\beta \mu$ and for varying asymmetry $\beta h = 0, \dots, 2.0$. 
Note that $\lambda > 0$ corresponds to attractive interactions, and interactions for $\lambda < 0$ are repulsive.
The insets show zooms into the region of positive $\beta \mu$, where quantum effects dominate.
We compare our CL results with third-order perturbation theory, imaginary-polarization calculations
(for the attractive case, as for repulsive interactions that option is not available), and the virial expansion 
in {the region $\beta \mu \leq -1.5$.}

The agreement between the methods is remarkable, in particular in the virial region (and for both
attractive and repulsive regimes), where except for very small deviations in the perturbative {third-order} answer, 
the results are almost indistinguishable from one another. Note that, although the virial coefficients
$b_2$ and $b_3$ used here vary considerably with the interaction strength $\lambda$ 
{(see Tab.~\ref{Table:VirialCoefficients}), the} dominant term at large negative $\beta \mu$ is interaction 
independent [cf. Eqs.~(\ref{Eq:LOVirialn}) and (\ref{Eq:LOVirialm})]; all the methods studied here reproduce 
that universal asymptotic behavior.
For $\beta \mu > 1$ the insets in Fig.~\ref{Fig:DensityLambda1} also show agreement of the CL 
results with the perturbative and iHMC numbers. 

Although the agreement between the various methods is remarkable, a word of caution is in order on the CL results 
for repulsive couplings. In that case, it was found in Ref.~\cite{Rammelmuller:2017vqn} that 
while the CL results for e.g., the ground-state energy, agree with the known exact results from the Bethe ansatz at zero temperature,
the distributions of the energy do not exhibit a finite variance 
(see also Refs.~\cite{PhysRevLett.107.201601, PhysRevD.86.014512, PhysRevB.92.125126, PhysRevE.93.043301, PhysRevE.93.033303}, where similar behavior is described in the context of cold atoms, QCD, entanglement, and 
electronic systems, even in the absence of a sign problem). 
This appears to be a general issue in QMC
studies and requires further investigation. In any case, the distributions in the attractive regime 
are statistically well-behaved.


\subsection{Polarization at $|\lambda|=1$}

\begin{figure*}[t]
	\centering
	\includegraphics[width=\columnwidth]{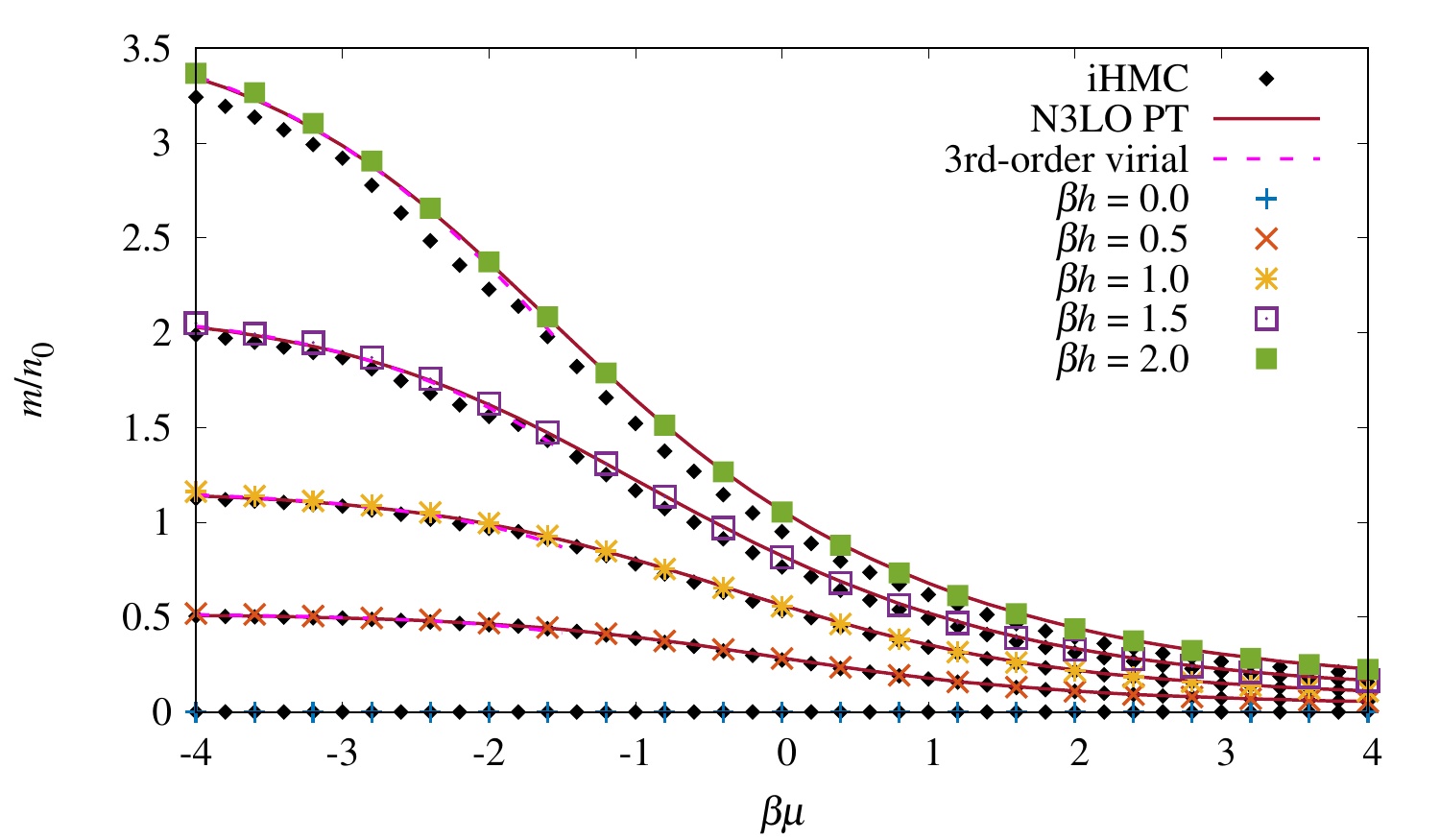}
	\includegraphics[width=\columnwidth]{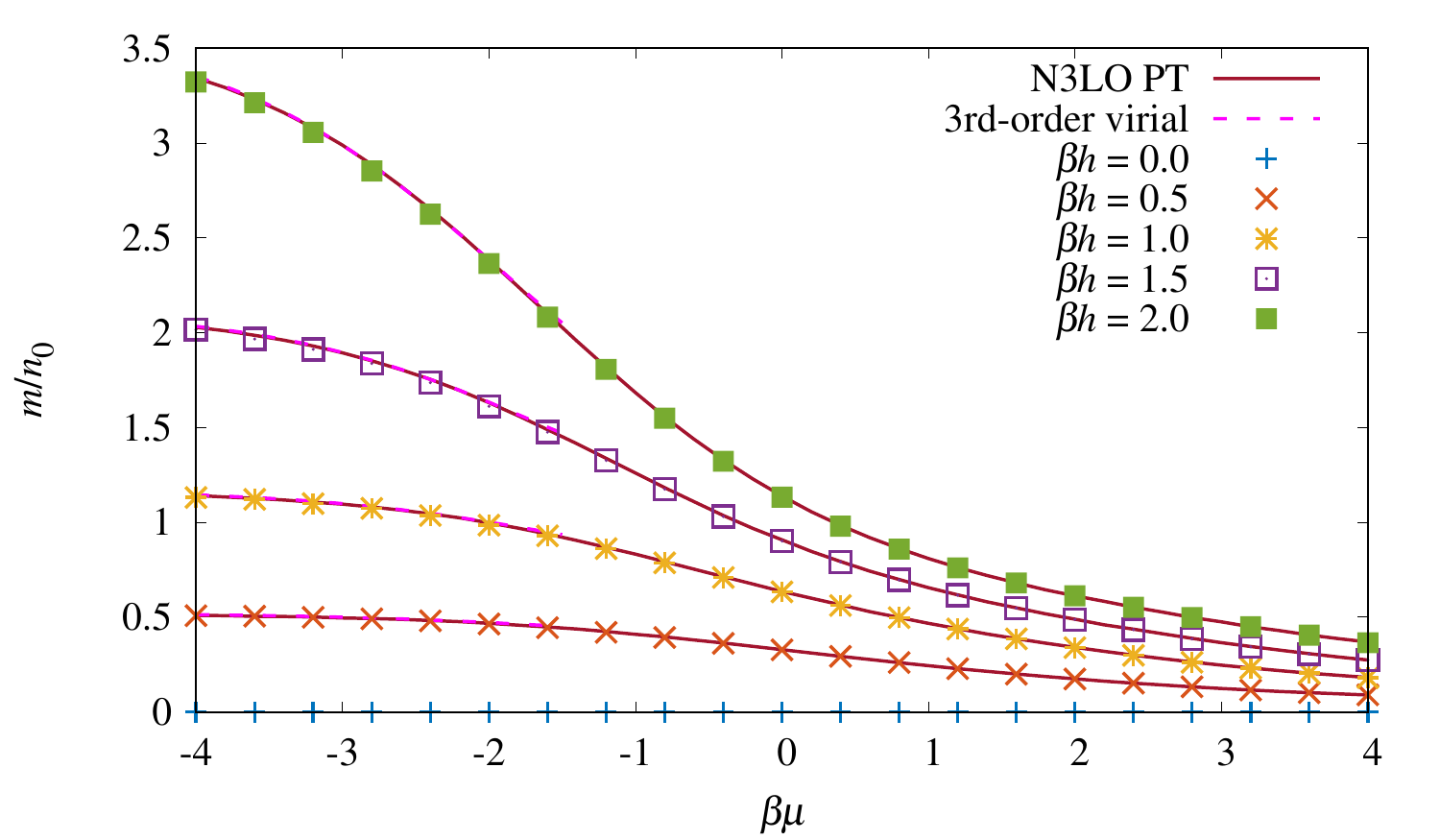}
	\caption{\label{Fig:MagnetizationLambda1}Spin polarization $m = n_\uparrow - n_\downarrow$ normalized by the non-interacting, unpolarized density $n_0$ for 
	{attractive (left) and repulsive (right) interactions of strength $\lambda = \pm 1$.}
	The CL results are shown with colored symbols, iHMC results (from Ref.~\cite{Loheac:2015fxa}) appear with
	black diamonds, perturbative results at third order are shown with solid lines, and virial expansion results
	appear as dashed lines.
	}
\end{figure*}

In Fig.~\ref{Fig:MagnetizationLambda1} we show our results for the polarization equation of 
state {at $\lambda = 1$ (left) and $\lambda = -1$ (right), 
as} a function of $\beta \mu$ and for varying asymmetry $\beta h = 0, \dots , 2.0$. 
Also in this case we compare our CL results with third-order perturbation theory, imaginary-polarization calculations
(for the attractive case), and the virial expansion {in the region $\beta \mu \leq -1.5$.
Once} again the results in the latter region are nearly indistinguishable from one another,
and they remain so for increasing $\beta \mu$ as well, as far as $\beta \mu = 4.0$ (where our
explorations concluded).
%

\subsection{Systematics of Langevin time discretization}

One of the features of stochastic quantization is that, either in its real or complex forms,
it performs a walk in configuration space with a specific fictitious time $t$ discretization,
which we denote here as $dt_\text{CL}$. Even when using adaptive algorithms, 
as done here, the adaptive-step tolerance effectively determines a scale for $dt_\text{CL}$ 
that affects the results. We have observed effects where if the tolerance is set such that it corresponds to an average time step which is too large, the CL evolution will converge to a value which systematically deviates from the true result. In addition, even for a fixed adaptive tolerance, a similar sensitivity exists for the initial $dt_\text{CL}$ used at the beginning of the trajectory.

\begin{figure}[t]
	\centering
	\includegraphics[width=\columnwidth]{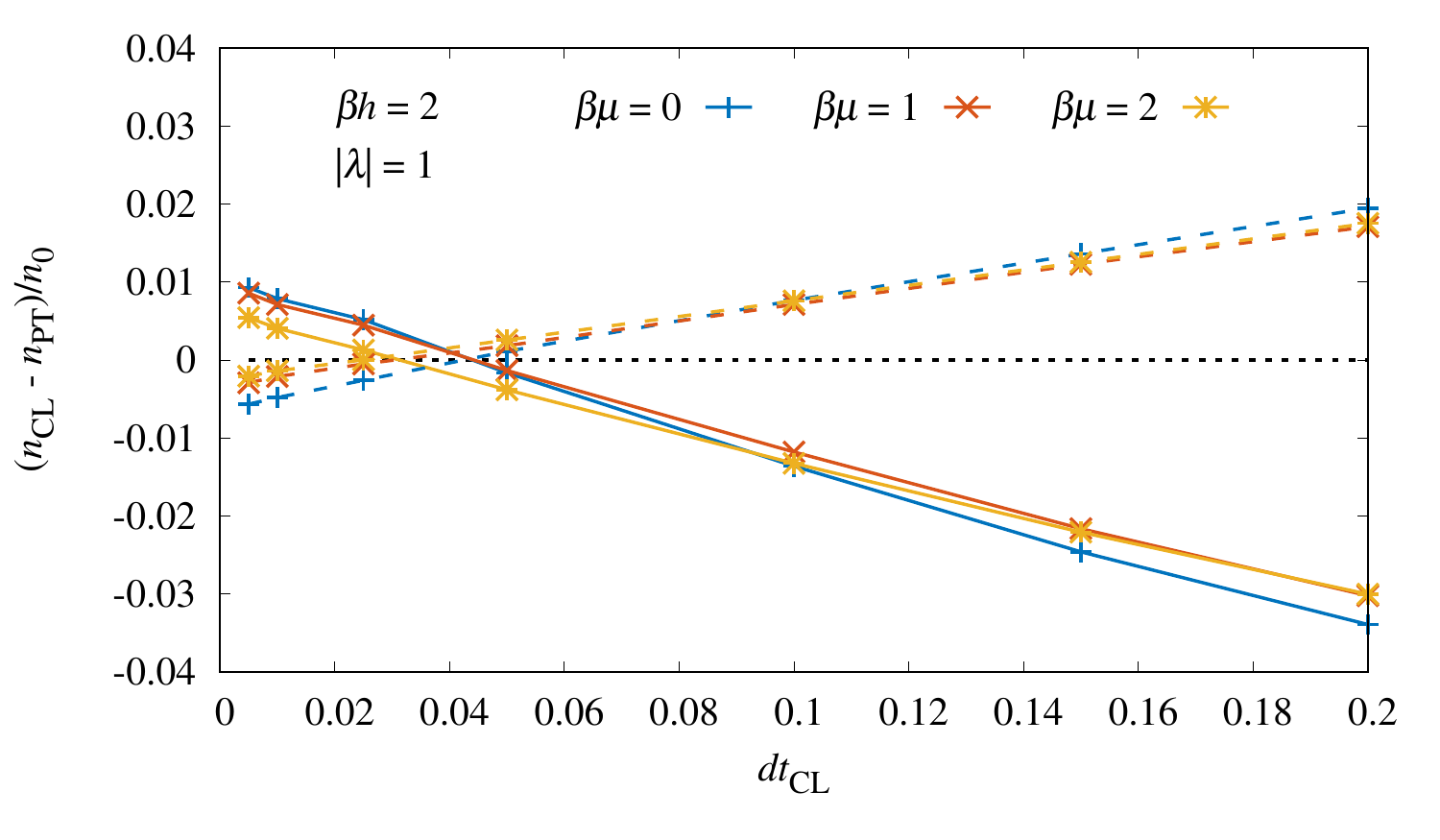}
	\caption{\label{Fig:TimestepSensitivity} Density equation of state $n = n_\uparrow + n_\downarrow$ 
	{at~$|\lambda|=1$ relative} to the third-order perturbative result $n^{}_\text{PT}$, normalized by the 
	non-interacting, unpolarized counterpart $n_0$, as a function of the CL time step $dt^{}_\text{CL}$, for three values of
	$\beta \mu$, all at $\beta h = 2.0$. The dashed horizontal line shows the $n = n^{}_\text{PT}$ line.
	The remaining differences at $dt_\text{CL} \to 0$ are likely a shortcoming of perturbation theory,
	but it would result in agreement between CL and the perturbative answers within the uncertainties
	shown in the other figures in this work, as in all cases the differences are reduced by a 
	factor of 3 when $dt_\text{CL} \to 0$.
	}
\end{figure}

To illustrate those effects, we show in Fig.~\ref{Fig:TimestepSensitivity} 
a plot of the sensitivity to the size of the initial CL time step $dt_\text{CL}$ {for~$|\lambda|=1$, using} 
the perturbative answer as a reference. 
As evident from that figure, the size of $dt_\text{CL}$  
is responsible for potential discrepancies. 
{The remaining difference between the CL and perturbative results in the limit $dt_\text{CL} \to 0$ 
is ascribed to the inaccuracy of N3LO perturbation theory.}
On the scale of the insets of Fig.~\ref{Fig:DensityLambda1}, however, that remaining difference would
appear as agreement between CL and perturbation theory. (The same {holds for the 
figures in} the Appendix.) {This highlights the} need to explore such systematic effects when using the CL method.

\section{Summary and Conclusions\label{Sec:Conclusions}}

In this work we have presented an application of the CL method to a classic problem:
the polarized one-dimensional Fermi gas. {With the main objective of validating the CL algorithm for non-relativistic Fermi gases, 
we compared our CL results for the finite-temperature density and polarization equations of state with those from 
perturbation theory, iHMC studies,
and the virial expansion.}

Generally speaking, our results speak favorably for the CL method as a way to tackle polarized matter, 
indicating that the door is open for calculations in higher dimensions and for non-trivial coupling
strengths.
More specifically, the results obtained with the various methods in the virial region are in remarkably good 
agreement with one another. {For~$\beta\mu \gtrsim -1.5$, small differences are noticeable in the
density equation of state at strong coupling ($\lambda = 2$), even less 
in the polarization.}

It should be pointed out that fermions in 1D can be addressed without a {sign problem 
by, e.g., mapping the system onto hard-core bosons (see e.g.~\cite{PhysRevB.26.5033}) or employing the fermion bag approach~\cite{Chandrasekharan:2013rpa}.
However, to our acknowledge, such methods do not generalize (efficiently) to higher dimensions,
which} is why we focused here {on benchmarks for auxiliary-field approaches. The} latter not only generalize to
higher dimensions but also to a wide range of situations including condensed matter, nuclear, and 
high-energy physics.

{{\it Acknowledgments.--} We thank C. R. Shill for providing values of the interacting virial coefficients 
and L.~Rammelm\"uller for} useful discussions.
This work was supported by HIC for FAIR within the LOEWE program of the State of Hesse
and by the National Science Foundation under Grants No.
DGE{1144081} (Graduate Research Fellowship Program),
PHY{1452635} (Computational Physics Program).


\appendix
\section{Results for higher interaction strengths: Density and polarization at $|\lambda| = 2$.}
\begin{figure*}[t]
	\centering
	\includegraphics[width=\columnwidth]{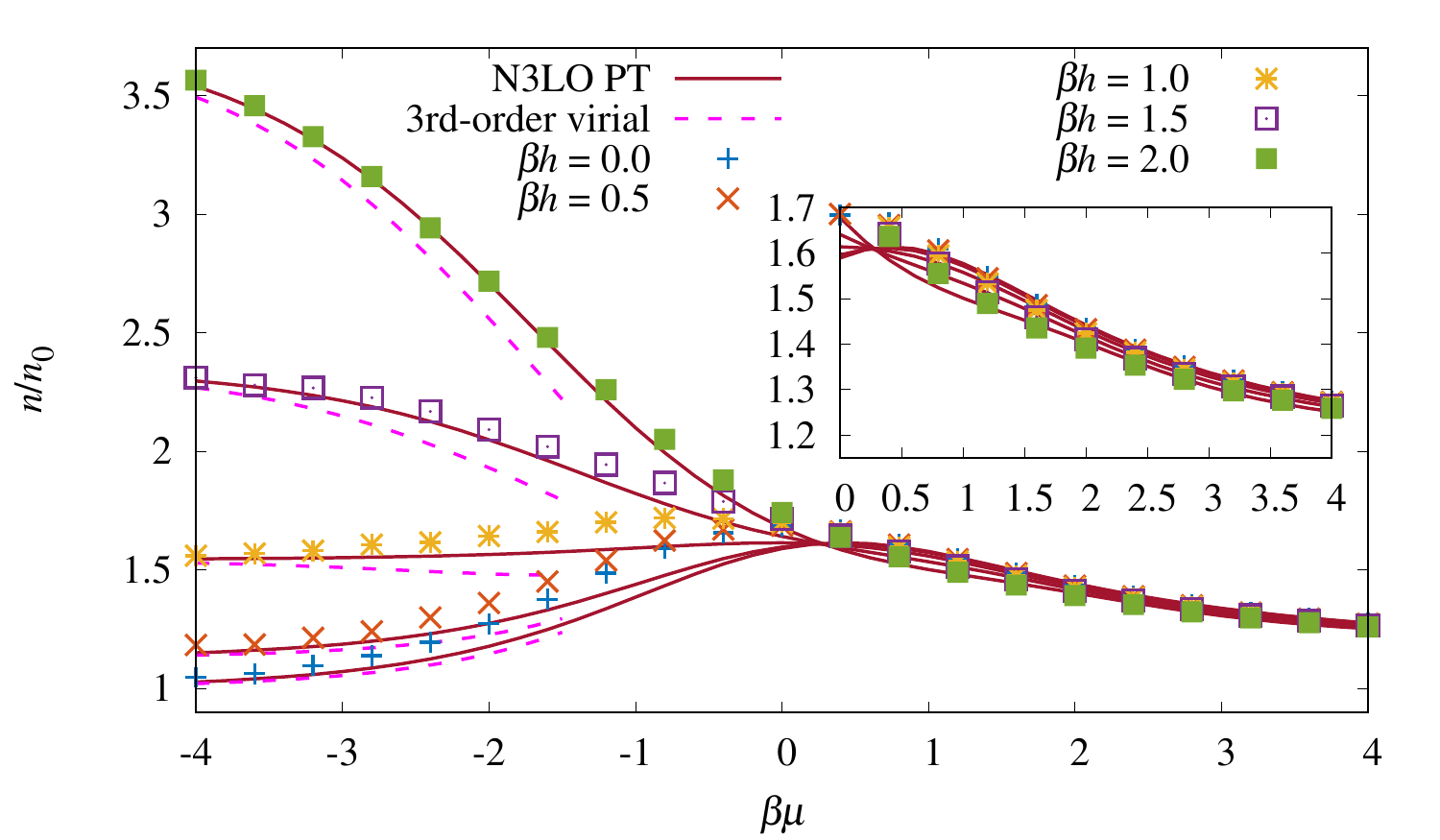}
	\includegraphics[width=\columnwidth]{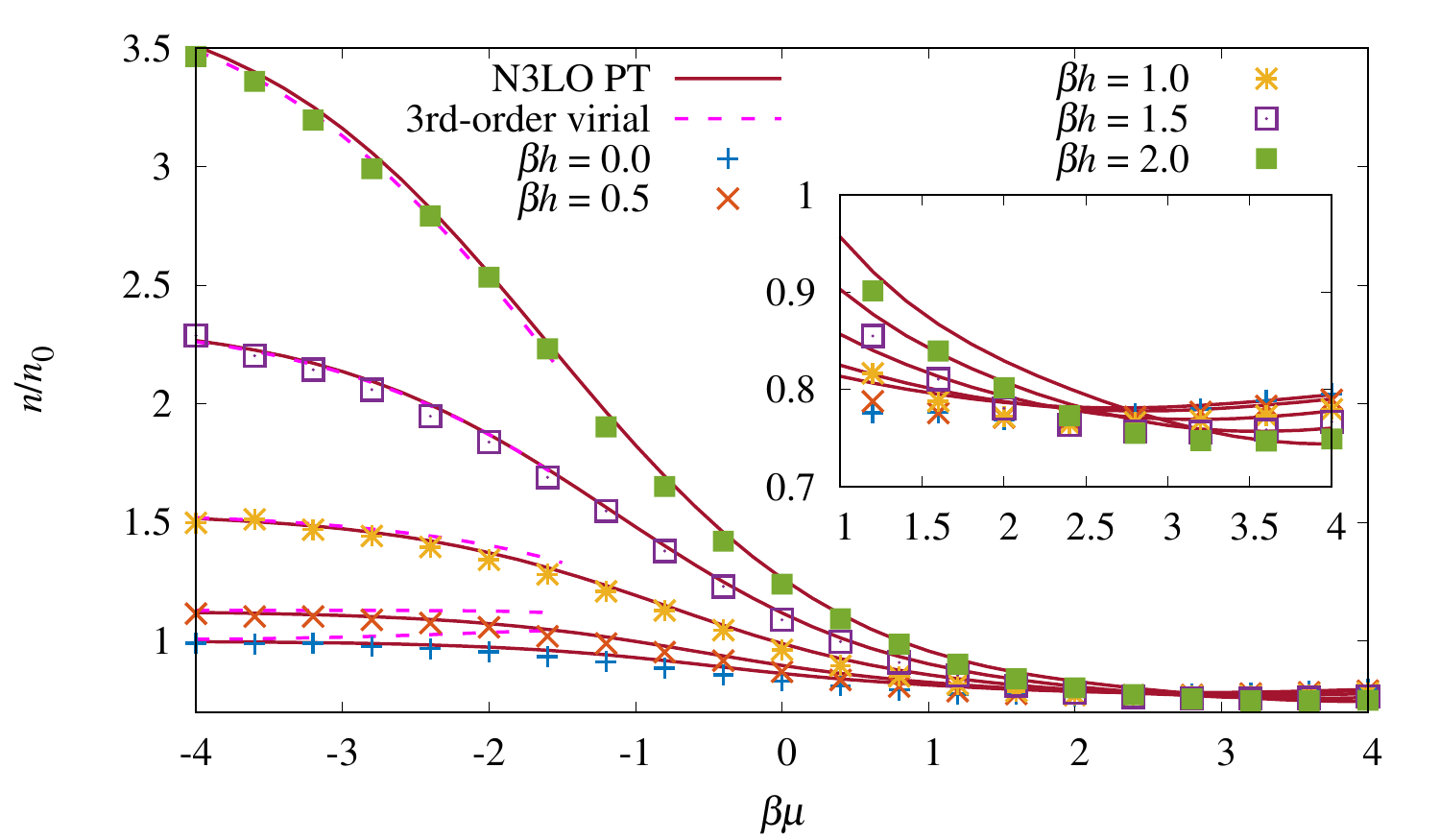}
	\caption{\label{Fig:DensityLambda2}Density equation of state $n = n_\uparrow + n_\downarrow$ normalized by the non-interacting, unpolarized 
	{counterpart $n_0$ for attractive (left) and repulsive (right) interactions of strength $\lambda = \pm 2$.
		Insets: Zoom in on the region $\beta\mu > 0$ (left) and $\beta\mu > 1$ (right).}
		The CL results are shown with colored symbols, perturbative results at third order are shown with solid lines, and virial expansion results appear as dashed lines.
	}
\end{figure*}

\begin{figure*}[t]
	\centering
	\includegraphics[width=\columnwidth]{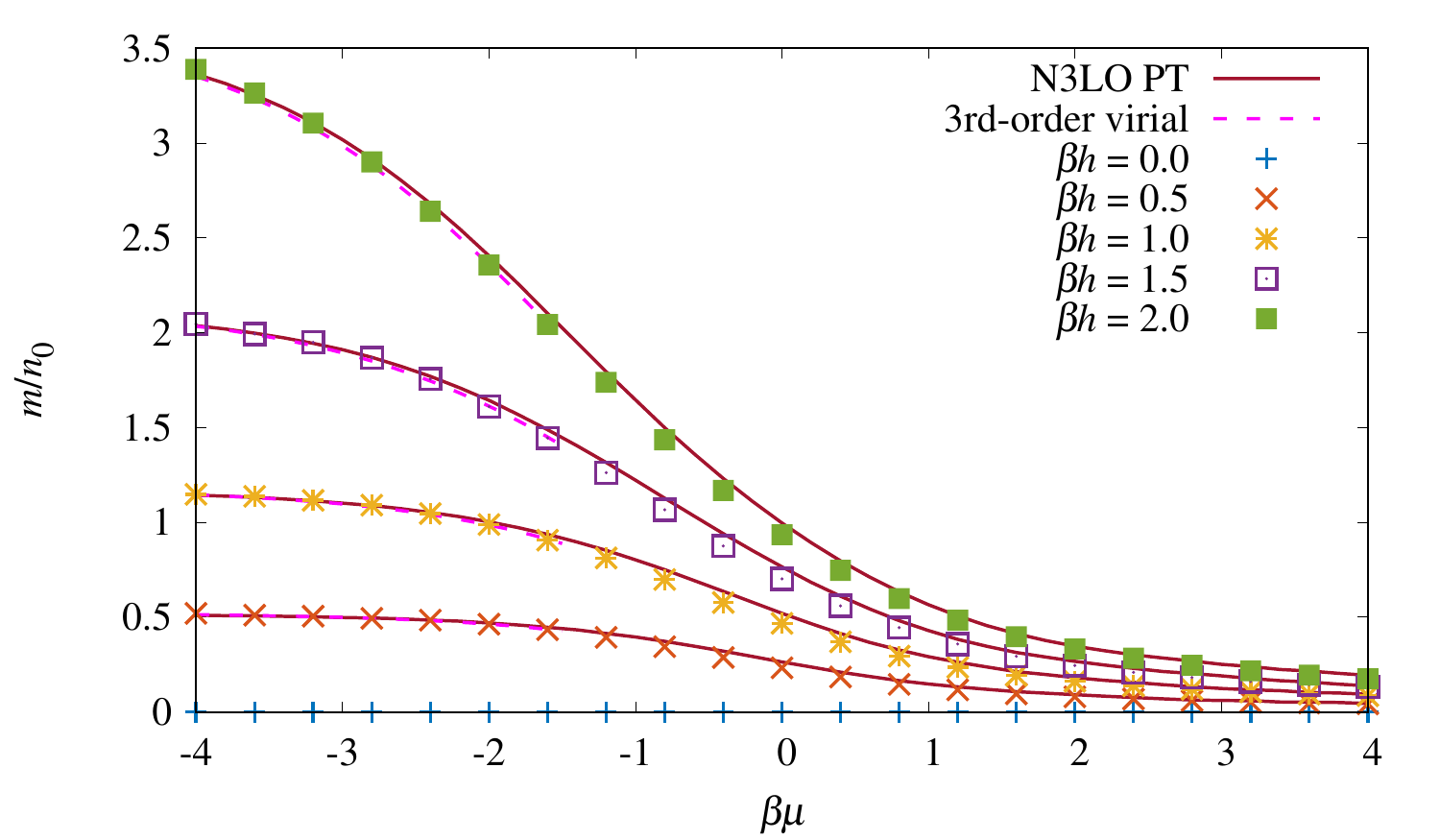}
	\includegraphics[width=\columnwidth]{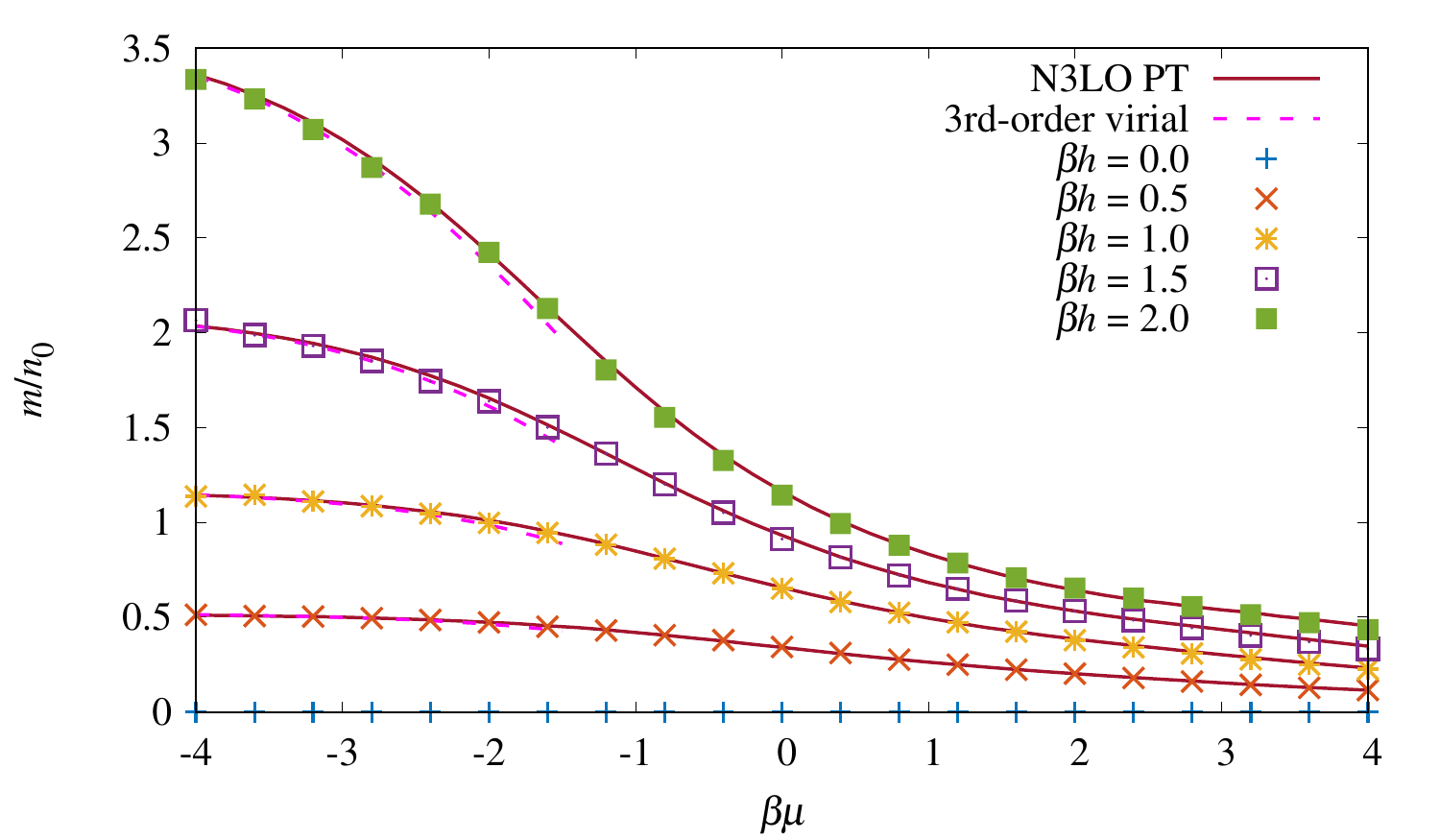}
	\caption{\label{Fig:MagnetizationLambda2}Spin polarization $m = n_\uparrow - n_\downarrow$ normalized by the non-interacting, unpolarized density $n_0$ 
	{for attractive (left) and repulsive (right) interactions} of strength $\lambda = \pm 2$. The CL results are shown with colored symbols, 
	perturbative results at third order are shown with solid lines, and virial expansion results appear as dashed lines.}
\end{figure*}

In this Appendix we present the density and polarization equations of state analogous to Figs.~\ref{Fig:DensityLambda1} and \ref{Fig:MagnetizationLambda1}, but for the stronger interaction strength of $|\lambda| = 2$. The same techniques discussed for the results at $|\lambda| = 1$ are applied here.

{In Fig.~\ref{Fig:DensityLambda2} we show our results for the density equation of state at $\lambda = 2$ (left) and 
$\lambda = -2$ (right), as} a function of $\beta \mu$ and for varying asymmetry $\beta h = 0, \dots, 2.0$. 
We compare our CL results with third-order perturbation theory and the virial expansion for $\beta \mu \leq -1.5$. 
As expected, the agreement between all three techniques deteriorates at the increased coupling strength when
compared to the {results for $|\lambda| = 1$. However, the overall comparison is satisfactory. For these systems, perturbation theory
is expected to break down at this coupling strength. Indeed, this is most obvious for the unpolarized case which
was further discussed in Ref.~\cite{Loheac:2017yar}. Agreement between perturbation theory and CL improves as the
polarization increases, where the effective interaction} between opposite spins lessens. The virial expansion demonstrates
a more significant deterioration at this coupling as both $\beta\mu$ and $\beta h$ move {away from $z_s \sim 0$.}

{In Fig.~\ref{Fig:MagnetizationLambda2}, we show our results for the polarization equation of state at $\lambda = 2$ (left) and 
$\lambda = -2$ (right), as} a function of $\beta \mu$ and for varying asymmetry $\beta h = 0, \dots , 2.0$. 
Also in this case we compare our CL results with perturbation theory calculations and find excellent
agreement for the whole range of $\beta \mu$ studied.

\section{Perturbative progression from first to third order.}
\begin{figure*}[t]
	\centering
	\includegraphics[width=\columnwidth]{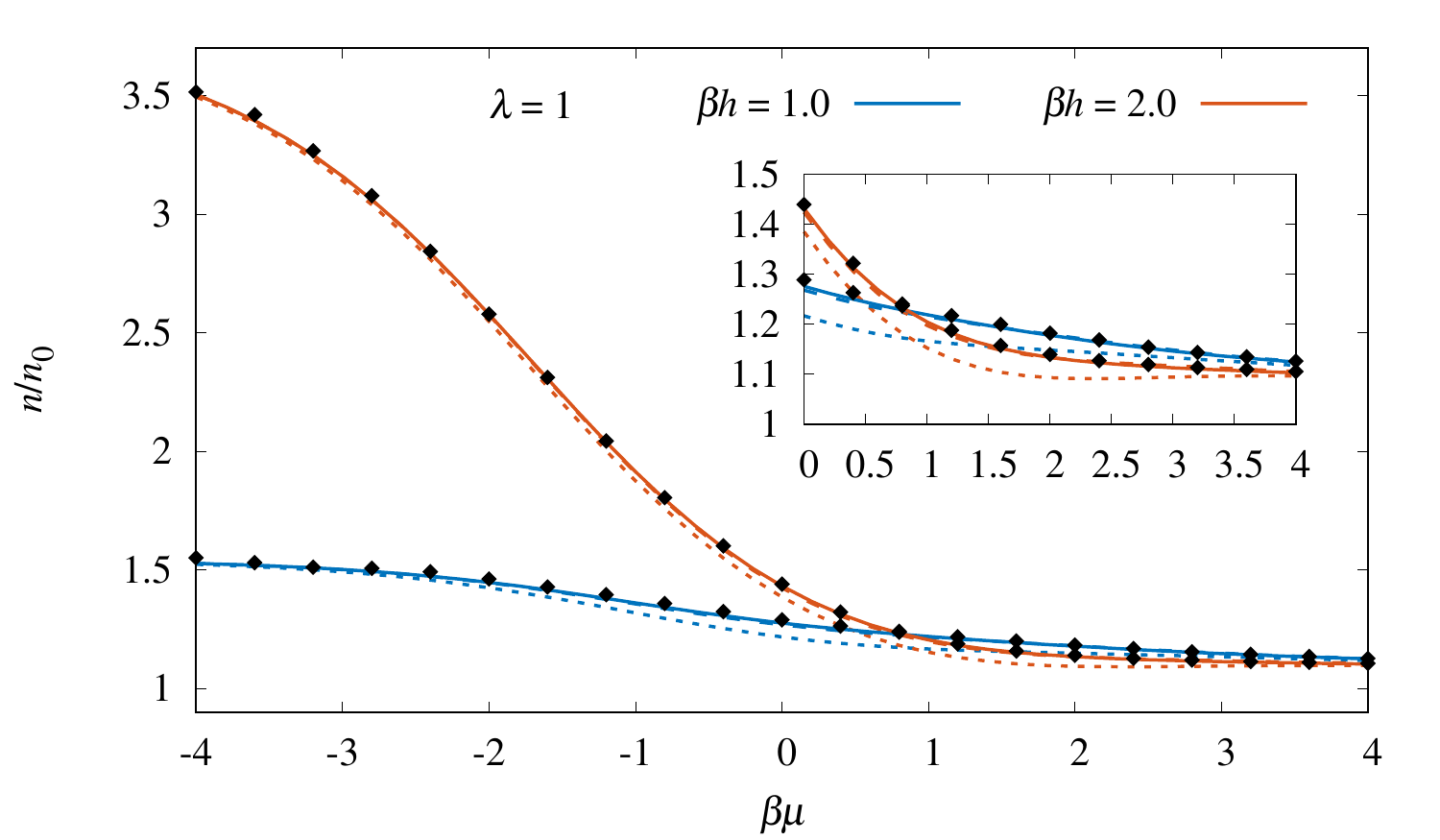}
	\includegraphics[width=\columnwidth]{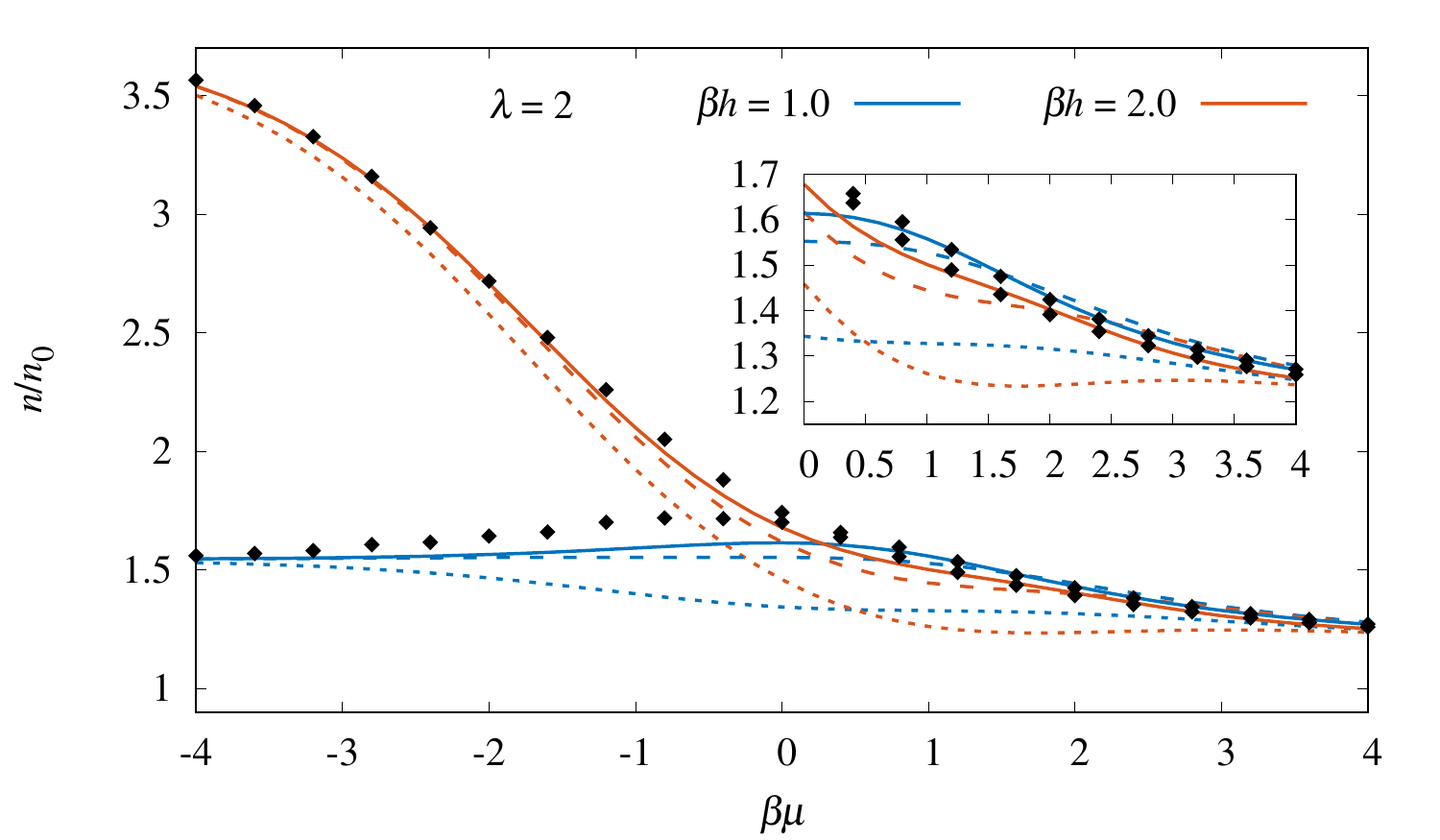}
	\caption{\label{Fig:DensityVsPT} Density $n = n_\uparrow + n_\downarrow$ normalized by the non-interacting, unpolarized counterpart $n_0$ 
	for attractive interaction of strength $\lambda = 1$ (left) and $\lambda = 2$ (right). 
	The CL results are shown with black diamonds, 
	perturbative results at first (NLO), second (N2LO), and third (N3LO) order are shown with dotted, dashed, and solid lines, respectively.}
\end{figure*}

Finally, in Fig.~\ref{Fig:DensityVsPT} we show the progression of density results in lattice perturbation theory at
first, second, and third order, for two attractive couplings ($\lambda = 1, 2$) and for two polarizations ($\beta h = 1.0, 2.0$). We note that the perturbative results appear very well converged at $\lambda = 1$, where they agree very well with the CL answers, as noted in the main text. On the other hand, at $\lambda = 2$, perturbation theory is (as expected) still slightly away from convergence [note in particular the big jump from first (dotted) to second order (dashed)], but it uniformly approaches the CL results.

\bibliography{bib}

\end{document}